\documentclass[aps, pra, superscriptaddress, twocolumn, nofootinbib]{revtex4-2}

\usepackage[english]{babel}
\usepackage{definitions}

\usepackage{amsmath}
\usepackage[colorlinks=true, allcolors=blue]{hyperref}
\hypersetup{breaklinks={true}}

\begin{document}

\title{Nonrelativistic spatiotemporal quantum reference frames}

\author{Michael Suleymanov}
\email{michael.suleymanov@biu.ac.il}
\affiliation{Faculty of Engineering and the Institute of Nanotechnology and Advanced Materials, Bar-Ilan University, Ramat Gan 5290002, Israel}

\author{Ismael L. Paiva}
\affiliation{H. H. Wills Physics Laboratory, University of Bristol, Tyndall Avenue, Bristol BS8 1TL, United Kingdom}

\author{Eliahu Cohen}
\affiliation{Faculty of Engineering and the Institute of Nanotechnology and Advanced Materials, Bar-Ilan University, Ramat Gan 5290002, Israel}

\begin{abstract}
Quantum reference frames have attracted renewed interest recently, as their exploration is relevant and instructive in many areas of quantum theory. Among the different types, position and time reference frames have captivated special attention. Here, we introduce and analyze a nonrelativistic framework in which each system contains an internal clock, in addition to its external (spatial) degree of freedom and, hence, can be used as a spatiotemporal quantum reference frame. We present expressions for expectation values and variances of relevant observables in different perspectives, as well as relations between these quantities in different perspectives in scenarios with no interactions. In particular, we show that even in these simple scenarios, the relative uncertainty between clocks affects the relative spatial spread of the systems.
\end{abstract}

\maketitle

\section{Introduction}

In nonrelativistic classical mechanics, space and time are absolute entities, existing independently of any physical object. Time serves as a parameter according to which a physical system evolves, and it is the same in every reference frame. Investigation of electrodynamics, followed by the formulation of Maxwell's equations, caused some friction with this notion. Maxwell's equations are invariant not under Galilean transformations but under Lorentz transformations, which do not keep time as a universal parameter. Taking the operational view that time is what is measured by clocks and clocks are also subject to the laws of physics, Einstein resolved the dispute between these two conflicting ideas by showing that mechanical systems should also transform according to Lorentz transformations. As a consequence, time is not the same in every reference frame.

In nonrelativistic quantum mechanics of a single particle, time serves as a Newtonian universal evolution parameter, while the spatial coordinate becomes an operator associated with a measurable quantity. Although the issue is more subtle in the case of multiple particles, in which the systems live in configuration space, this asymmetry signals a tension with the theory of special relativity, where time and space are treated on equal footing in Minkowski spacetime. This tension arises from the lack of invariance properties within the algebra of operators~\cite{hilgevoord2002time, hilgevoord2005time}. Relativistic versions of quantum mechanics (Klein-Gordon and Dirac), where the number of described particles, and, hence the number of degrees of freedom, is fixed, suffer from various problems and inconsistencies. So far, the most successful and powerful theory that combines both principles of quantum mechanics and special relativity is relativistic quantum field theory (RQFT). In RQFT, space and time coordinates are parameters and fields become operators describing an infinite number of degrees of freedom, while particles are field excitations.

Moreover, in nonrelativistic quantum mechanics, there have been discussions about the introduction of a time observable since the early days of the theory, including the known objection by Pauli~\cite{pauli1933allgemeinen}, which can be resolved with the use of positive operator-valued measures (POVMs)~\cite{garrison1970canonically, busch1995operational}. The possibility of such an operator was discussed in terms of Heisenberg cut in Ref.~\cite{aharonov1961time}. Alternative approaches related to these discussions attempt to combine quantum mechanics with special relativity, e.g., the study of relativistic quantum dynamics, introduced by Steuckelberg in 1941~\cite{Stueckelberg:1941rg} and further developed after that~\cite{Horwitz:1973ak, Fanchi:1978jt, Horwitz:2015mka}. In these approaches, time becomes an operator associated with a measurable quantity, called \textit{coordinate time}, just like space coordinates. However, in addition, these approaches introduce an extra invariant and universal time parameter, which is sometimes called \textit{historical} or \textit{universal} time and serves as an evolution parameter like the one in nonrelativistic classical and quantum mechanics. Using this framework, several systems were described, e.g., relativistic Coulomb-like and harmonic oscillator potentials~\cite{Arshansky_Horwitz1989} and relativistic spacetime string~\cite{Suleymanov:2017wlp}.

In this work, the relevant approach for time is the Page and Wootters framework~\cite{PW83}. Differently from Steuckelberg's approach, the Page and Wootters framework introduces relational dynamics from the constraint given by the Wheeler-DeWitt equation with the use of clock systems. This framework has gained considerable attention lately~\cite{giovannetti2015quantum, marletto2017evolution, smith2019quantizing, diaz2019history, hohn2020switch, Castro_Ruiz_2020, smith2020quantum, carmo2021quantifying, ballesteros2021group, Trinity, paiva2022flow, baumann2022noncausal, paiva2022dynamical, adlam2022watching, paiva2022non, altaie2022time, rijavec2023robustness} in the larger context of quantum reference frames~\cite{Aharonov_Susskind_ssr, Aharonov_Kaufherr_qrf, Rovelli_1991, Bartlett2007, angelo2011physics, pereira2015galilei, loveridge2017relativity, Giacomini_2019, martinelli2019quantifying, 2020_change_of_perspective, Yang2020switchingquantum, de2021perspective, castro2021relative}, where even the notion of a subsystem may depend on the choice of reference frame~\cite{ahmad2022quantum}. We combine it with the concept of spatial quantum reference frames (SQRFs). Specifically, we consider a system of $N$ particles, each with an internal clock degree of freedom. This allows us to study the expression for expectation values and variances of relevant observables in a given perspective as well as relations between these quantities in different perspectives. As a particular consequence of this analysis, we show that relative uncertainty between clocks generally affects the spatial distribution of systems in a given perspective.

The remainder of this work is structured as follows. In Section~\ref{sec:pre} we briefly review the basics of spatial and temporal reference frames. With this, we introduce our framework in Section~\ref{sec:STQRF}, following up with an analytic study of expectation values and variances of relevant observables in Section~\ref{sec:ExpVal}. Next, in Section~\ref{sec:examples} we present some consequences implied by the analysis in Section~\ref{sec:ExpVal}. Finally, we conclude our work and present some prospective ideas in Section~\ref{sec:disc}.

\section{Preliminaries}
\label{sec:pre}

The fundamental concept behind this work is the notion of a frame of reference. In studies related to it, we typically have a number of systems, some of which can be used as a reference frame to describe the physics of the others. Before any further consideration is made, the Hilbert space $\mathcal{H}^{\text{kin}}$, called the kinematical Hilbert space, can be constructed as a tensor product of the Hilbert space associated with each individual system.

However, at this stage, a reference frame has not been chosen yet, and the system must satisfy a constraint (or a list thereof). We will consider in this work the case with first-class constraints. Then, not every vector in $\mathcal{H}^{\text{kin}}$ will represent a possible physical state~\cite{dirac1964lectures}. This is the case because, in the presence of a constraint $\hat{C}$, we are interested in a subset of vectors $\ket{\Psi}^{\text{phys}}$ of the total space satisfying $\hat{C} \ket{\Psi}^{\text{phys}} = 0$. The resultant space of states that satisfy the constraint is the \textit{physical} Hilbert space $\mathcal{H}^{\text{phys}}$. Using the technique known as group averaging~\cite{ashtekar1995quantization, giulini1999generality, giulini2000group, marolf2002group, thiemann2007modern}, observe that, given $\ket{\Psi}^{\text{kin}} \in \mathcal{H}^{\text{kin}}$, the vector $\delta(\hat{C}) \ket{\Psi}^{\text{kin}}$, where
\begin{equation}
    \delta(\hat{C}) \coloneqq \frac{1}{2\pi} \int_{G_C} ds \,\, e^{i s\hat{C}},
\end{equation}
satisfies the constraint $\hat{C}$ and, if no further constraints apply, belongs to $\mathcal{H}^{\text{phys}}$. In the above expression, $G_C$ refers to a set of values that depend on the constraint $\hat{C}$.

Within $\mathcal{H}^{\text{phys}}$, there still exists a multiplicity of descriptions of the systems since a frame of reference has not been chosen yet. It turns out that this is a type of gauge freedom, and choosing a frame is equivalent to fixing a gauge \cite{henneaux1992quantization, 2020_change_of_perspective}.

In this work, we combine spatial and temporal quantum reference frames. To better organize the presentation and make the discussion less abstract, we first briefly review some basics of each of them.

\subsection{Spatial quantum reference frames}
\label{subsec:SQRFs}

Consider a system of $3$ particles, each with position $\hat{x}_I$ and momentum $\hat{p}_I$, where $I\in\mathfrak{I}$ and $\mathfrak{I}\coloneqq\{A,B,C\}$. The kinematic space associated with the joint system is $\mathcal{H}^{\text{kin}} = \mathcal{H}_A \otimes \mathcal{H}_B \otimes \mathcal{H}_C$. Since the idea is to use any of the particles as a position reference frame, it is, in a sense, natural to impose the momentum constraint \cite{2020_change_of_perspective}
\begin{equation}
    \hat{P}_T\ket{\Psi}^{\text{phys}} = 0,
    \label{eq:p-constraint}
\end{equation}
where $\hat{P}_T \coloneqq \sum_{I\in\mathfrak{I}} \hat{p}_I$. The inclusion of this constraint implies that the coordinates $\hat{x}_I$ do not have a meaning by themselves. Instead, only their relative values with respect to each other have a meaning.

\begin{figure}
	\centering
	\includegraphics[scale=0.85]{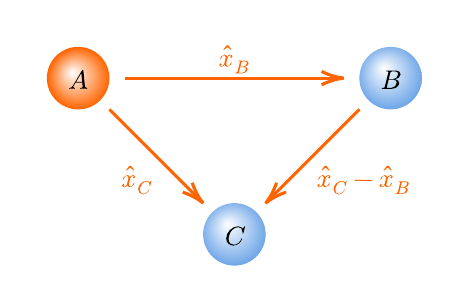}
    \caption{Illustration of a spatial perspective-dependent description when particle $A$ is chosen to be a reference frame.}
	\label{FIG:SQRF_A}
\end{figure}

An arbitrary state in $\mathcal{H}^{\text{kin}}$ can be written as
\begin{equation}
    \begin{aligned}
        \ket{\Psi}^{\text{kin}} &= \int dp_A dp_B dp_C \Psi^\text{kin}(p_A,p_B,p_C) \ket{p_A, p_B, p_C} \\
           &= \int d\underline{P} \Psi^\text{kin}(\underline{P}) \ket{\underline{P}},
    \end{aligned}
\end{equation}
where we have introduced $\underline{P} \coloneqq \{p_I; \; I \in \mathfrak{I}\}$ and $d\underline P=\prod_{I\in\mathfrak{I}}dp_I$. However, because of the constraint in Eq.~\eqref{eq:p-constraint}, a physical state is of the form
\begin{subequations}
    \begin{align}
        \ket{\Psi}^{\text{phys}} &=\int d\underline{p_{\bar{A}}} \psi_{\bar{A}}(\underline{p_{\bar{A}}}) \ket{-p_{\bar{A}}, \underline{p_{\bar{A}}}}_{ABC} \label{eq:a-persp-general} \\
        &=\int d\underline{p_{\bar{B}}} 
        \psi_{\bar B}(\underline{p_{\bar{B}}}) \ket{-p_{\bar{B}}, \underline{p_{\bar{B}}}}_{BAC} \\
        &=\int d\underline{p_{\bar{C}}} 
        \psi_{\bar C}(\underline{p_{\bar{C}}}) \ket{-p_{\bar{C}}, \underline{p_{\bar{C}}}}_{CAB},
    \end{align}
    \label{eq:wf-space-neutral}
\end{subequations}
where $p_{\bar{S}} \coloneqq \sum_{I\in\mathfrak{I}\setminus\{S\}} p_I$ for every $S\in\mathfrak{I}$, $\underline{p_{\bar{S}}} \coloneqq \underline{P} \setminus \{p_S\}$, and
\begin{equation}
    \begin{aligned}
        \psi_{\bar A}(\underline{p_{\bar{A}}}) &\coloneqq 
        \Psi^\text{kin}(-p_{\bar{A}}, p_B, p_C), \\
        \psi_{\bar B}(\underline{p_{\bar{B}}}) &\coloneqq \Psi^\text{kin}(p_A, -p_{\bar{B}}, p_C), \\
        \psi_{\bar C}(\underline{p_{\bar{C}}}) &\coloneqq \Psi^\text{kin}(p_A, p_B, -p_{\bar{C}}).
    \end{aligned}
\end{equation}
or, in general,
\begin{equation}
\psi_{\bar I}(\underline{p_{\bar I}})=
\int dp_I\delta(P_T)\Psi^\text{kin}(\underline{P}).
\label{eq:psi_bar_I}
\end{equation}
Each of the three expressions in Eq.~\eqref{eq:wf-space-neutral} includes a degree of freedom that is now redundant, which is associated with expressing $\ket{\Psi}^{\text{phys}}$ from the perspective of different particles. In case one of these perspectives is chosen, say, $A$'s perspective in Eq.~\eqref{eq:a-persp-general}, this redundancy can be removed with a projection onto $\ket{x_A=0}$~\cite{2020_change_of_perspective}, as illustrated in Fig.~\ref{FIG:SQRF_A}. The resulting reduced state $\ket{\psi_{\bar A}} \coloneqq \sqrt{2\pi} \braket{x_A=0 | \Psi}^{\text{phys}}$ is
\begin{equation}
    \ket{\psi_{\bar A}}=
    \int d\underline{p_{\bar{A}}}
    \psi_{\bar A}(\underline{p_{\bar{A}}}) \ket{\underline{p_{\bar{A}}}},
    \label{eq:reduced-state-a}
\end{equation}
which belongs to $\mathcal{H}_B \otimes \mathcal{H}_C$, can be constructed. It is noteworthy that $x_A$ is a fixed parameter that can be arbitrarily chosen. The choice of $x_A=0$ is generally made for convenience.

The inner product between $\ket{\Psi}^{\text{phys}}$ and $\ket{\Phi}^{\text{phys}}$ can be defined as~\cite{henneaux1992quantization, giulini2000group, marolf2002group, 2020_change_of_perspective}
\begin{equation}
    \left(\Psi^{\text{phys}},\Phi^{\text{phys}}\right)_{\text{phys}} \coloneqq \prescript{\text{kin}}{}{\bra{\Psi}}\de(\hat{P}_T)\ket{\Phi}^{\text{kin}}
\end{equation}
This construction is consistent with the usual requirement of having a normalized wave function (in a given frame) and, moreover, is independent of the choice of perspective. Indeed, in momentum representation, the physical inner product can be written as
\begin{equation}
    \begin{aligned}
    (\Psi^{\text{phys}},\Phi^{\text{phys}})_{\text{phys}} &= \int d\underline{p_{\bar{A}}} 
    \psi_{\bar A}^*(\underline{p_{\bar{A}}})
    \phi_{\bar A}(\underline{p_{\bar{A}}}) \\
    &=\int d\underline{p_{\bar{B}}} 
    \psi_{\bar B}^*(\underline{p_{\bar{B}}})
    \phi_{\bar B}(\underline{p_{\bar{B}}}) \\
    &=\int d\underline{p_{\bar{C}}}
    \psi_{\bar C}^*(\underline{p_{\bar{C}}})
    \phi_{\bar C}(\underline{p_{\bar{C}}}).
    \end{aligned}
\end{equation}
Now, suppose the joint system composed of $A$, $B$, and $C$ has a dynamics governed by the Hamiltonian
\begin{equation}
    \hat H_T= \sum_{I\in\mathfrak{I}} \frac{\hat p_I^2}{2m_I} + \sum_{I,J\in\mathfrak{I};I>J} V_{IJ}(\hat{x}_I - \hat{x}_J).
\end{equation}
In $A$'s reference frame, it becomes
\begin{equation}
\hat{H}_{\bar A} \coloneqq \frac{1}{2} \sum_{I \neq A} 
\frac{\hat{p}_I^2}{m_I}  + \frac{\(\hat{p}_B+ \hat{p}_C\)^2}{2m_A}  + V(\hat{x}_B,\hat{x}_C),
\end{equation}
where $V(\hat{x}_B,\hat{x}_C) \coloneqq V_{BA}(\hat{x}_B) + V_{CA}(\hat{x}_C) + V_{CB}(\hat{x}_C - \hat{x}_B)$. In fact, it can be shown~\cite{2020_change_of_perspective} that $\hat{H}_{\bar A}$ governs the dynamics of $\ket{\psi_{\bar A}}$, i.e.,
\begin{equation}
    i\frac{d}{dt}\ket{\psi_{\bar A}(t)} = \hat{H}_{\bar A} \ket{\psi_{\bar A}(t)},
\end{equation}
where $\ket{\psi_{\bar A}(0)} = \ket{\psi_{\bar A}}$.

\subsection{Temporal quantum reference frames: Page--Wootters framework}\label{subsec:TQRFs}

The framework introduced by Page and Wootters was proposed as a solution to a time issue that arises in the canonical quantization of general relativistic systems~\cite{PW83}. Specifically, the quantization leads to the constraint
\begin{equation}
    \hat{H}_T\ket{\Psi}^{\text{phys}}=0,
    \label{eq:wdw}
\end{equation}
known as Wheeler-DeWitt equation~\cite{dewitt1967quantum}. This constraint implies that the total system (supposedly the entire universe) is an energy eigenstate and, hence, does not evolve in time. A question then surfaces: How to reconcile this with the standard notion of time evolution?

Approaching this issue, Page and Wootters showed that the standard dynamics could indeed emerge from this static scenario in a relational way if the entire system was split into two parts: The first, associated with $\mathcal{H}_C$, is a clock and the other, associated with $\mathcal{H}_R$, is the system whose dynamics will be analyzed. Moreover, they assumed that $\hat{H}_T = \hat{w} + \hat{H}_R$, where $\hat{w}$ is the Hamiltonian of the clock and $\hat{H}_R$ is the Hamiltonian of system $R$.

In $\mathcal{H}_C$, time states are defined as
\begin{equation}
    |t\rangle \coloneqq e^{-i\hat{w}(t-t')}|t'\rangle
\end{equation}
with the condition that $\int_G dt \; |t\rangle\langle t| = \mathds{1}_C$, where the set $G$ refers to the values necessary to parametrize the one-dimensional group generated by $\hat{\omega}$~\cite{Trinity}. In this expression, as in this entire work, we use units such that $\hbar=1$. With this, it is possible to define the time operator
\begin{equation}
    \hat{t} \coloneqq \int_G dt \; t \; |t\rangle\langle t|,
    \label{eq:t-def}
\end{equation}
and it holds that $[\hat{t},\hat{\omega}] = i\mathds{1}_C$ and $\frac{d}{dt} |t\rangle = -i \hat{w} |t\rangle$~\cite{Trinity, paiva2022non}. Observe that $\hat{t}$ and $\hat{\omega}$ are not necessarily canonically conjugate, i.e., a Heisenberg pair. This is the case because such pairs are composed of Hermitian operators, but $\hat{t}$ is only guaranteed to be symmetric. In fact, $\hat{t}$ is Hermitian only if $\langle t|t'\rangle = \delta(t-t')$ for every $t$ and $t'$, in which case the clock is referred to as an ideal clock. When this condition is not matched, $\hat{t}$ is the first moment of a positive operator-valued measure~\cite{smith2019quantizing, Trinity}.

While ideal clocks do not overcome Pauli's objection (they require the clock's Hamiltonian to be unbounded from below), they can be seen as approximations for clock systems with large spectra. Moreover, their use is convenient since they simplify mathematical derivations. For this reason, we assume ideal clocks throughout this work. With this choice, we also omit $G$ from the integral over $t$ in Eq.~\eqref{eq:t-def} since, in this case, we have an integral over the entire real line.

With this set and defining $|\psi(t)\rangle \coloneqq \sqrt{2\pi} \langle t| \Psi\rangle^{\text{phys}}$, Eq.~\eqref{eq:wdw} implies that $\langle t| \hat{H}_T|\Psi\rangle^{\text{phys}} = 0$, which leads to
\begin{equation}
    i\frac{d}{dt}\ket{\psi(t)} = \hat{H}_R\ket{\psi(t)},
\end{equation}
which is the standard Schr\"odinger equation. The factor $\sqrt{2\pi}$ is typically not introduced in the definition of $|\psi(t)\rangle$, and has been included here for symmetry with the spatial reference frames.

The inner product between $\ket{\Psi}^{\text{phys}}$ and $\ket{\Phi}^{\text{phys}}$ is defined as~\cite{henneaux1992quantization, giulini2000group, marolf2002group, smith2019quantizing, Trinity}
\begin{equation}
    \begin{aligned}
\left(\Psi^{\text{phys}},\Phi^{\text{phys}}\right)_{\text{phys}} \coloneqq& \prescript{\text{kin}}{}{\bra{\Psi}}\de(\hat H_T)\ket{\Phi}^{\text{kin}} \\
        =& \braket{\psi(t)|\phi(t)},
    \end{aligned}
\end{equation}
where $\ket{\phi(t)} \coloneqq \sqrt{2\pi} \braket{t|\Phi}^{\text{phys}}$

It is noteworthy that, while the original formulation of the framework assumed no interactions with the clock, this framework has been extended to the interacting case~\cite{smith2019quantizing}. Moreover, system $R$ may be a multipartite system and even include multiple clocks, each of which could be a potential reference frame for time~\cite{hohn2020switch, Castro_Ruiz_2020, paiva2022flow, paiva2022dynamical, paiva2022non}.

\section{$1+1$ spatiotemporal quantum reference frames}
\label{sec:STQRF}

We now combine the two frameworks presented in the previous section. More precisely, we consider a composite system of three particles, each with degrees of freedom associated with time (i.e., a clock) and space, with their respective Hilbert spaces $\mathcal{H}_{C_I}$ and $\mathcal{H}_{R_I}$, where $I\in\mathfrak{I}$. Each particle $I$ has external degrees of freedom associated with the position operator $\hat{x}_I$ and its canonical conjugate momentum $\hat{p}_I$. Moreover, the internal clock of each particle $I$ has a Hamiltonian $\hat{\omega}_I$ and a time operator $\hat{t}_I$, built according to Page and Wootters' description. The full Hilbert space under consideration is, then, $\mathcal{H}^{\text{kin}} \equiv \bigotimes_I \left(\mathcal{H}_{C_I}\otimes \mathcal{H}_{R_I}\right)$, and an arbitrary element $\ket{\Psi}^{\text{kin}}\in\mathcal{H}^{\text{kin}}$ is
\begin{equation}
    \ket{\Psi}^{\text{kin}}=\int d\underline{\Omega} \; d\underline{P} \; \ket{\underline{\Omega},\underline{P}} \Psi^{\text{kin}}(\underline{\Omega},\underline{P}),
    \label{eq:general-kin}
\end{equation}
where $\underline{\Omega} \coloneqq \{\omega_I; I\in\mathfrak{I}\}$.

We assume the entire joint system satisfies the momentum constraint in Eq. \eqref{eq:p-constraint} and the Hamiltonian constraint in Eq. \eqref{eq:wdw}. In the latter, $\hat{H}_T$ denotes the total Hamiltonian of the system, assumed to be of the form
\begin{equation}
    \hat{H}_T \coloneqq \sum_{I\in\mathfrak{I}} 
    \left(\hat{w}_I + \frac{\hat{p}_I^2}{2m_I}\right).
\end{equation}
Hence, $[\hat P_T,\hat H_T]$ vanishes everywhere in $\mathcal{H}^{\text{kin}}$. As a consequence, $\delta(\hat{P}_T)$ and $\delta(\hat{H}_T)$ also commute\footnote{This would also hold if the Hamiltonian was of the form
\begin{equation*}
    \hat{H}_T \coloneqq \sum_{I\in\mathfrak{I}} \left(\hat{w}_I + \frac{\hat{p}_I^2}{2m_I}\right) + \sum_{I,J\in\mathfrak{I};I>J} V_{IJ}(\hat{x}_I - \hat{x}_J).
\end{equation*}
However, for the argument to be more direct, we consider the noninteracting case.}. Then, unequivocally, we can consider the map from $\mathcal{H}^{\text{kin}}$ into $\mathcal{H}^{\text{phys}}$
\begin{equation}
    \ket{\Psi}^{\text{kin}} \mapsto \ket{\Psi}^{\text{phys}} \coloneqq \de(\hat P_T) \de(\hat H_T) \ket{\Psi}^{\text{kin}}.
    \label{eq:def-phys-state}
\end{equation}
Then, we can write, in general,
\begin{equation}
    \begin{aligned}
&\ket{\Psi}^{\text{phys}}=
\int d\underline{\Omega}d\underline{P}
\ket{\underline{\Omega},\underline{P}}
\delta(H_T)\delta(P_T)
\Psi^\text{kin}(\underline{\Omega},\underline{P})
    \end{aligned}
\end{equation}
or, in perspective-dependent views,
\begin{equation}
    \begin{aligned}
        &\ket{\Psi}^{\text{phys}}=\\
        &=\int d\underline{\omega_{\bar{A}}} d\underline{p_{\bar{A}}} \ket{\omega_A = -H_{\bar A}, \; \underline{\omega_{\bar{A}}}, \; p_A=-p_{\bar A}, \; \underline{p_{\bar{A}}}} \\
        &\hspace{9mm} \Psi^{\text{kin}}(-H_{\bar A},\om_B,\om_C,-p_{\bar A},p_B,p_C) \\
        &=\int d\underline{\omega_{\bar{B}}} d\underline{p_{\bar{B}}} \ket{\omega_B = -H_{\bar B}, \; \underline{\omega_{\bar{B}}}, \; p_B=-p_{\bar B}, \; \underline{p_{\bar{B}}}} \\
        &\hspace{9mm} \Psi^{\text{kin}}(\om_A,-H_{\bar B},\om_C,p_A,-p_{\bar B},p_C)\\
        &=\int d\underline{\omega_{\bar{C}}} d\underline{p_{\bar{C}}} \ket{\omega_C = -H_{\bar C}, \; \underline{\omega_{\bar{C}}}, \; p_C=-p_{\bar C}, \; \underline{p_{\bar{C}}}} \\
        &\hspace{9mm} \Psi^{\text{kin}}(\om_A,\om_B,-H_{\bar C},p_A,p_B,-p_{\bar C}),
    \end{aligned}
    \label{Psi^phys}
\end{equation}
where $\underline{\omega_{\bar{I}}} \coloneqq \underline{\Omega} \setminus \{\omega_I\}$ and $H_{\bar{I}}$ is the ``scalar version'' of the operator
\begin{equation}
    \hat{H}_{\bar{I}} = \sum_{J\in\mathfrak{I}\setminus\{I\}} \hat{\omega}_J + \frac{\hat p_{\bar{I}}^2}{2m_I} + 
    \sum_{J\in\mathfrak{I}\setminus\{I\}} \frac{\hat{p}_J^2}{2m_J} = \sum_{J\in\mathfrak{I}\setminus\{I\}} \hat{\omega}_J+\hat{\mathcal{K}}_{\bar I},
    \label{H_bar}
\end{equation}
where $\hat{\mathcal{K}}_{\bar I}\equiv \hat p_{\bar{I}}^2/2m_I + \sum_{J\in\mathfrak{I}\setminus\{I\}} \hat{p}_J^2/2m_J$. As will be seen soon, this is the effective Hamiltonian of the systems from $I$'s perspective. By scalar version, we mean the eigenvalue of the operator associated with the eigenstate $\ket{\underline{\omega_{\bar{A}}}, \underline{p_{\bar{A}}}}$.

Denoting 
\begin{equation}
\Psi_{\bar I}=
\int d\omega_I dp_I\delta(\Omega_T)\delta(P_T)\Psi^\text{kin}(\underline{\Omega},\underline{P}),
\label{eq:Psi_bar_I}
\end{equation}
or, explicitly, for the systems $A$, $B$, and $C$,
\begin{equation}
    \begin{aligned}
        &\Psi_{\bar A} \equiv \Psi^{\text{kin}}\left(-H_{\bar A},\om_B,\om_C,-p_{\bar A},p_B,p_C\right),\\
        &\Psi_{\bar B} \equiv \Psi^{\text{kin}}(\om_A,-H_{\bar B},\om_C,p_A,-p_{\bar B},p_C),\\
        &\Psi_{\bar C} \equiv \Psi^{\text{kin}}(\om_A,\om_B,-H_{\bar C},p_A,p_B,-p_{\bar C}),
    \end{aligned}
    \label{Psi_J}
\end{equation}
we observe that $\Psi_{\bar A}$, $\Psi_{\bar B}$, and $\Psi_{\bar C}$ describe the same physical state in $\mathcal{H}^{\text{phys}}$. However, they correspond to different perspectives before the removal of the redundant degrees of freedom.

\begin{figure}[t]
	\centering
	\includegraphics[scale=0.85]{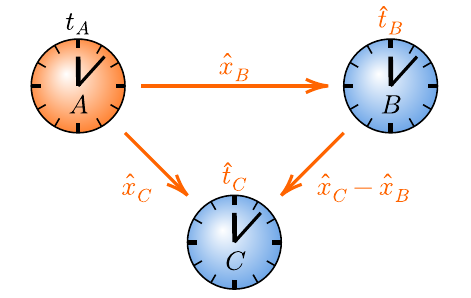}
	\caption{Perspective in a spatiotemporal reference frame. In addition to their spatial degree of freedom, each particle is assumed to have an internal degree of freedom identified as a clock. Here, system $A$ is chosen to be a reference frame. In this description, $\hat{x}_B$, $\hat{x}_C$, $\hat{t}_B$, and $\hat{t}_C$ are observables, while $t_A$ is a parameter.}
	\label{FIG_STQRF_A}
\end{figure}

With this, we follow the procedure discussed in the previous section and define the state of the system conditioned on the spatiotemporal state of one of the particles, say, $A$, which will serve as a reference frame as a ``ruler'' and a ``clock'' (illustrated in Fig.~\ref{FIG_STQRF_A}). Such a reduced state is
\begin{equation}
    \ket{\psi_{\bar A}(t_A)} \coloneqq 2\pi \Braket{t_A,x_A=0|\Psi}^{\text{phys}},
    \label{eq:def-a-perspective}
\end{equation}
where $\psi_{\bar{A}}$ represents the state of systems $B$ and $C$ from $A$'s perspective. In this picture, $\hat{x}_{I}$, $\hat{p}_{I}$, $\hat{t}_{I}$ and $\hat{\omega}_{I}$ are observables, where $I\in\mathfrak{I}\setminus\{A\}$, while $t_A$ is a parameter of evolution.

In this definition, while $t_A$ is a quantity to be varied for the study of the time evolution of the systems from the perspective of $A$, $x_A=0$ is a fixed parameter used as a reference to give relational meaning to the position of the other particles.

Applying the constraint in Eq. \eqref{eq:wdw} and projecting it onto $\ket{t_A, x_A=0}$, we obtain the Schr\"odinger equation
\begin{equation}
    i\frac{d}{dt_A}\ket{\psi_{\bar A}(t_A)} = \hat{H}_{\bar{A}}\ket{\psi_{\bar A}(t_A)}.
    \label{eq:a-dynamics}
\end{equation}
Observe that, from the initial conditions, we have
\begin{equation}
    \begin{aligned}
        &\ket{\psi_{\bar A}(t_A)} =
        e^{-i t_A \hat{H}_{\bar{A}}} 
        \ket{\psi(0)} \\
             &= 2\pi \bra{t_A=0,x_A=0} 
             e^{-it_A \hat{H}_{\bar{A}}} \ket{\Psi}^{\text{phys}}.
    \end{aligned}\label{A_evolution}
\end{equation}
Then, from Eq.~\eqref{Psi^phys}, we obtain
\begin{equation}
    \begin{aligned}
        |\psi_{\bar{A}}(t_A)\rangle &= \int d\underline{\omega_{\bar{A}}} \; d\underline{p_{\bar{A}}} \ket{\underline{\omega_{\bar{A}}}, \underline{p_{\bar{A}}}} e^{-i t_AH_{\bar{A}}} 
        \Psi_{\bar{A}} \\
        &= \int d\underline{\omega_{\bar{A}}} \; d\underline{p_{\bar{A}}} \ket{\underline{\omega_{\bar{A}}}, \underline{p_{\bar{A}}}} \psi_{\bar{A}}(t_A, \underline{\omega_{\bar{A}}}, \underline{p_{\bar{A}}}),
    \end{aligned}
    \label{A_pers}
\end{equation}
where we have introduced
\begin{equation}
    \psi_{\bar{A}}(t_A, \underline{\omega_{\bar{A}}}, \underline{p_{\bar{A}}}) \coloneqq \langle \underline{\omega_{\bar{A}}}, \underline{p_{\bar{A}}} | \psi_{\bar{A}}(t_A)\rangle = e^{-it_A H_{\bar{A}}} \Psi_{\bar{A}}.
    \label{eq:psi-a-bar}
\end{equation}
Observe that, from the definition in Eqs.~\eqref{eq:def-phys-state} and \eqref{eq:def-a-perspective}, applying the constraints on a kinematical state and reducing the resulting state to a perspective is equivalent to solving the Schr\"odinger equation in Eq.~\eqref{eq:a-dynamics}.

Repeating the above steps, one may obtain a description with respect to any subsystem. For example, the state from {$B$'s perspective} is of the form
\begin{equation}
    \begin{aligned}
        |\psi_{\bar B}(t_B)\rangle = \int &d\underline{\omega_{\bar{B}}} d\underline{p_{\bar{B}}} \ket{\underline{\omega_{\bar{B}}},\underline{p_{\bar{B}}}} \psi_{\bar{B}}(t_B, \underline{\omega_{\bar{B}}}, \underline{p_{\bar{B}}})
    \end{aligned}
    \label{B_pers}
\end{equation}
with
\begin{equation}
    \psi_{\bar{B}}(t_B, \underline{\omega_{\bar{B}}}, \underline{p_{\bar{B}}}) = e^{-it_B H_{\bar{B}}} \Psi_{\bar{B}},
    \label{eq:psi-b-bar}
\end{equation}
which obeys the following Schr\"odinger equation
\begin{equation}
    i\frac{d}{dt_B} 
    \ket{\psi_{\bar{B}}(t_B)} = \hat H_{\bar{B}} \ket{\psi_{\bar{B}}(t_B)}.
\end{equation}

Based on previous approaches derived from the coherent group averaging technique~\cite{henneaux1992quantization, ashtekar1995quantization, giulini1999generality, giulini2000group, marolf2002group, Trinity}, we introduce the inner product
\begin{equation}
    \left(\Psi^{\text{phys}}, \Phi^{\text{phys}}\right)_{\text{phys}} \coloneqq \prescript{\text{kin}}{}{\Bra{\Psi}} \delta(\hat{H}_T) \delta(\hat{P}_T) \Ket{\Phi}^{\text{kin}}.
\label{inner_product_st}
\end{equation}
Observe that $(\Psi^{\text{phys}}, \Phi^{\text{phys}})_{\text{phys}} = \Braket{\psi(t_I) | \phi(t_I)}$ for every $I\in\mathfrak{I}$, i.e., the normalization of reduced states is independent of perspective.

\section{Expectation values, variances and covariances}\label{sec:ExpVal}

We now explore the perspective-dependent expressions describing expectation values and variances of the relevant operators in the framework just introduced. Different from the simplification considered in the previous section, we will assume a general number of subsystems, i.e., $\mathfrak{I}\coloneqq\{A,B,C,\hdots\}$.

Recall that the expression for the expectation value of an operator $\hat O_{\bar I}$ (that does not depend explicitly on time) from the perspective of the system $I$ at an instant of time $t_I$ is
\begin{equation}
    \braket{\hat{O}_{\bar I}}_I(t_I) = 
    \bra{\psi_{\bar I}(t_I)}
    \hat{O}_{\bar I}
    \ket{\psi_{\bar I}(t_I)}.
\end{equation}
Similarly, its variance is
\begin{equation}
    \sigma^2(\hat{O}_{\bar I})_{I}(t_I)=
    \braket{\hat{O}_{\bar I}^2}_I(t_I) -
    \braket{\hat{O}_{\bar I}}_I^2(t_I).
\end{equation}
When it is sufficiently clear, we omit the dependency on $t_I$.

Moreover, for any integrable function $f(\underline{\Omega},\underline{P})$ and for every $I,J\in\mathfrak{I}$, it holds that
\begin{equation}
    \label{eq:identity-any-frame}
    \begin{aligned}
        \int d\underline{\Omega} d\underline{P} &\delta(H_T)\delta(P_T) f(\underline{\Omega},\underline{P}) \\
        &= \int d\underline{\omega_{\bar I}}d\underline{p_{\bar I}} \; f(\omega_I=-H_{\bar I},\underline{\omega_{\bar I}},p_I=-p_{\bar I},\underline{p_{\bar I}}) \\
        &=\int d_{\bar I} f_{\bar I}=\int d_{\bar J} f_{\bar J},
    \end{aligned}
\end{equation}
where $f_{\bar{I}} \equiv f(\omega_I=-H_{\bar I},\underline{\omega_{\bar I}},p_I=-p_{\bar I},\underline{p_{\bar I}})$ and $d_{\bar I}\equiv d\underline{\omega_{\bar{I}}} d\underline{p_{\bar{I}}}$.

\subsection{Momentum and energy}

We start by observing that the expectation value of the momentum of subsystem $I$ is time-independent and independent of the reference frame. Indeed, $(e^{it_J H_{\bar{J}}}\Psi_{\bar J}^*) p_I (e^{-it_J H_{\bar{J}}} \Psi_{\bar J}) = \Psi_{\bar J}^*p_I\Psi_{\bar J}$ and then, for every $t_J\in G$,
\begin{equation}
    \<\hat{p}_I\>_{J} = \int d_{\bar J} \Psi_{\bar J}^*p_I\Psi_{\bar J} = \int d_{\bar K}\Psi_{\bar K}^*p_I\Psi_{\bar K}= \<\hat{p}_I\>_{K},
    \label{eq:<p_I>_any_short}
\end{equation}
where $J,K\in\mathfrak{I}\setminus\{I\}$. 
Moreover, using Eq.~\eqref{eq:identity-any-frame}, we have
\begin{equation}
    \begin{aligned}
        \<\hat p_I\>_{J} &= \int d\underline{\Omega} \; d\underline{P} {\Psi^{\text{kin}}}^*(\underline{\Omega}, \underline{{P}}) \delta(P_T)\delta(H_T) p_I {\Psi^{\text{kin}}}(\underline{\Omega}, \underline{{P}}) \\
        &= \int d_{\bar{I}} \Psi^*_{\bar{I}} (-p_{\bar{I}}) \Psi_{\bar{I}} \\
        &= -\braket{\hat{p}_{\bar{I}}}_{I}=
        -\sum_{K\in\mathfrak{I}\setminus\{I\}}\braket{\hat{p}_K}_I.
        \label{eq:<p_I_J=-sum}
    \end{aligned}
\end{equation}

Furthermore, using the notion of covariance between operators
\begin{equation}
    \text{cov}(\hat{A},\hat{B})\coloneqq \braket{\hat{A}\hat{B}} - \braket{\hat{A}}\braket{\hat{B}},
\end{equation}
we conclude, as shown in Appendix~\ref{app:p-omega}, that the covariance between $\hat{p}_I$ and $\hat{p}_J$ (for $I\neq J$) in an arbitrary reference frame $M\in\mathfrak{I}\setminus\{I,J\}$ is
\begin{equation}
    \text{cov}(\hat{p}_I,\hat{p}_J)_M = -\sigma^2(\hat{p}_I)_{J} - \sum_{L\in\mathfrak{I}\setminus\{I,J\}}\text{cov}(\hat{p}_I,\hat{p}_L)_{J},
    \label{eq:sigma^2(p_I)_J}
\end{equation}
i.e., it can be computed in terms of quantities associated with the reference frame of system $J$ (or of system $I$) alone.

Similarly, for the clocks' energies, the expectation value of $I$'s clock energy from $J$'s perspective satisfies
\begin{equation}
    \braket{\hat{\omega}_I}_J = \int d_{\bar{J}} \Psi_{\bar{J}}^* \omega_I \Psi_{\bar{J}} = \int d_{\bar{K}} \Psi_{\bar{K}}^* \omega_I \Psi_{\bar{K}} = \braket{\hat{\omega}_I}_K.
    \label{eq:<omega_I>_J}
\end{equation}
Moreover, as we show in Appendix~\ref{app:p-omega}, the covariance between $\hat{\omega}_I$ and $\hat{\omega}_J$ (for $I\neq J$) in an arbitrary perspective $M\in\mathfrak{I}\setminus\{I,J\}$ can also be given in terms of quantities associated with the reference frame of system $J$ (or system $I$) alone. More precisely,
\begin{equation}
    \begin{aligned}
        \text{cov}(\hat{\omega}_I,\hat{\omega}_J)_M &= -\sigma^2(\hat{\omega}_I)_J - 
        \sum_{L\in\mathfrak{I}\setminus\{I,J\}} \text{cov}(\hat{\omega}_I,\hat{\omega}_L)_{J} \\
        &\quad -\text{cov}(\hat{\omega}_I,\hat{\mathcal{K}}_{\bar J})_J.
        \label{eq:sigma(omega_I)_J}
    \end{aligned}
\end{equation}

\subsection{Time}
\label{sec:time}

Since there is no interaction between the systems and the clocks or even among different clocks, one might expect that the ``flow of time'' is the same in every clock. This idea can be indeed validated by the fact that
\begin{equation}
    \begin{aligned}
        \braket{\hat{t}_I}_{J}(t_J) &=\braket{\psi_{\bar{J}}(t_J)|\hat{t}_I|\psi_{\bar J}(t_J)} \\
        &= \int d_{\bar{J}} \Psi_{\bar{J}}^* e^{i t_J H_{\bar{J}}} i\frac{d}{d\omega_I}\left(e^{-i t_J H_{\bar J}} \Psi_{\bar{J}}\right) \\
        &= \int d_{\bar{J}} \Psi_{\bar J}^* \left(i\frac{d}{d\omega_I} + t_J\right) \Psi_{\bar J} \\
        &= \braket{\hat{t}_I}_{J}(t_J=0) + t_J.
    \end{aligned}
    \label{<t_I>_J_short}
\end{equation}
Moreover, as shown in Appendix~\ref{Appendix:Time},
\begin{equation}
    \sigma^2\left(\hat{t}_I\right)_J(t_J) = \sigma^2\left(\hat{t}_I\right)_J(t_J=0),
    \label{eq:var-sigma-t}
\end{equation}
i.e., the variance of a clock in a given frame is time-independent. This means that besides an initial offset between the clock's average value, their main difference is associated with their initial variance.

If the initial average value and variance of a clock $I$ in the reference frame $J$ are known, according to Eq.~\eqref{eq:app:<f(t_I)>_J}, the reciprocal quantities are also known.
The reciprocal temporal expectation values are opposite
\begin{equation}
    \begin{aligned}
        \braket{\hat{t}_I}_{J}(t_J=0) 
        &= -\braket{\hat{t}_J}_{I}(t_I=0)
    \end{aligned}
    \label{<t_I>_J(0)}
\end{equation}
and the variances coincide,
\begin{equation}
    \sigma^2\left(\hat{t}_I\right)_J(t_J)
    = \sigma^2\left(\hat{t}_J\right)_I(t_I),
    \label{eq:time-var-recip}
\end{equation}
as proved in Appendix~\ref{Appendix:Time} and illustrated in Fig.~\ref{FIG:reciprocal-time}.

\begin{figure}[t]
	\centering
	\includegraphics[scale=0.85]{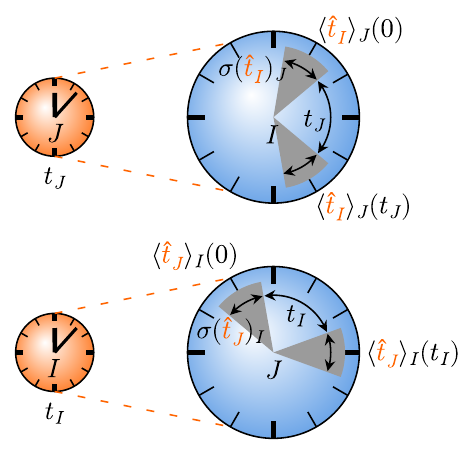}
	\caption{Representation of reciprocal temporal descriptions. In general, the clocks are not synchronized but have the same variance in each others' reference frames.}
	\label{FIG:reciprocal-time}
\end{figure}

\subsection{Velocity}

Before considering expectation values and variances of systems' positions in different reference frames, it is convenient to introduce the (noncanonical) velocity operator. Based on the standard notion that velocity is the time derivative of the position, the velocity $\hat{v}_{I\vert J}$ of a particle $I$ in the reference frame of system $J$ is introduced as~\cite{mandelstam1945uncertainty, messiah1961quantum, aharonov1975measurement, aharonov2005quantum}
\begin{equation}
    \hat{v}_{I \vert J} \coloneqq i [\hat{H}_{\bar{J}}, \hat{x}_I] = 
    \frac{\hat{p}_I}{m_I} + \frac{\hat{p}_{\bar{J}}}{m_J}.
    \label{v_I(J)}
\end{equation}
This coincides with the definition of relative velocity between $I$ and $J$ in the kinematical space. Indeed, the latter corresponds to
\begin{equation}
    \hat{v}_{IJ} \coloneqq \frac{\hat{p}_I}{m_I} - \frac{\hat{p}_J}{m_J}.
\end{equation}
With this, it can be checked that $\left(\hat{v}_{IJ}\right)_{\bar J} \equiv \hat{v}_{I \vert J}$ or, equivalently, $\left(\hat{v}_{JI}\right)_{\bar I} \equiv \hat{v}_{J \vert I}$. Since it follows from the definition that $\hat{v}_{IJ} = -\hat{v}_{JI}$, it holds that
\begin{equation}
    \braket{\hat{v}_{I\vert J}}_J = -\braket{\hat{v}_{J\vert I}}_I,
    \label{<v_{I|J>_J}}
\end{equation}
which corresponds to an expected property of reciprocal velocities.

\subsection{Position}
\label{sec:position}

We are now ready to study a few properties of expectation values and variances of position operators in different frames. Details of the calculations presented here can be found in Appendix~\ref{Appendix:Space}.

To start, observe that the expectation value of $\hat x_I$ in $J$'s reference frame for some $I\neq J$ satisfies
\begin{equation}
    \begin{aligned}
        \braket{\hat{x}_I}_J(t_J) &= \bra{\psi_{\bar{J}}(t_J)} \hat{x}_I \ket{\psi_{\bar{J}}(t_J)} \\
        &= \int d_{\bar{J}} \Psi_{\bar{J}}^* e^{it_J H_{\bar{J}}} i\frac{d}{dp_I} e^{-it_JH_{\bar J}} \Psi_{\bar J} \\
        &= \braket{\hat{x}_I}_J(t_J=0) + \braket{\hat{v}_{I \vert J}}_Jt_J.
    \end{aligned}
    \label{eq:x_I(t_J)}
\end{equation}
Then, on average, the expression for the position of particle $I$ is the same given by a free nonrelativistic classical particle. Deviations from the classical case can be seen from the time evolution of the variance of $\hat{x}_I$, which is
\begin{equation}
    \begin{aligned}
        \sigma^2(\hat{x}_I)_J(t_J) &= \sigma^2(\hat{x}_I)_J(t_J=0) - t_J\widetilde{\text{cov}}(\hat{x}_I,\hat{v}_{I\vert J})_J(t_J=0) \\
        &\quad + t_J^2\sigma^2(\hat{v}_{I\vert J}^2)_J(t_J=0),
        \label{eq:sigma(x_I)_J}
    \end{aligned}
\end{equation}
where
\begin{equation}
    \widetilde{\text{cov}}(\hat{A}, \hat{B}) \coloneqq \frac{1}{2} (\braket{\hat{A}\hat{B} + \hat{B}\hat{A}}) - \braket{\hat{A}}\braket{\hat{B}}.
    \label{eq:cov-tilde}
\end{equation}

\begin{figure}[t]
	\centering 
	\includegraphics[scale=0.65]{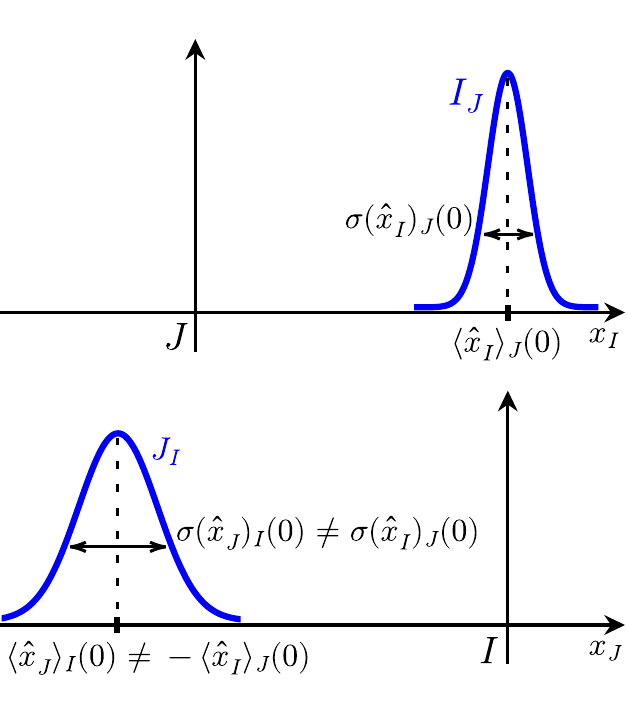}
	\caption{Representation of reciprocal spatial descriptions. Due to the different reference frames for time associated with each system, there is generally a shift that violates the anti-symmetric position expectation value and, moreover, the reciprocal spatial variance is not the same.}
	\label{FIG:sigma_position}
\end{figure}

When considering the reciprocal of expectation values, the symmetry observed in Eqs.~\eqref{<t_I>_J(0)} and \eqref{eq:time-var-recip} do not hold. Indeed, it can be shown that the relation between the initial expectation value of position in a given frame and its reciprocal, according to Eq.~\eqref{app:<x_I>_J(t_J)}, is
\begin{equation}
    \braket{\hat{x}_I}_J(t_J=0) = -\braket{\hat{x}_J}_I(t_I=0) +\braket{\hat{v}_{J \vert I} \hat{t}_J}_I(t_I=0).
    \label{<x_I>J}
\end{equation}
Finally, we can write the following for the initial variance of a clock and its reciprocal
\begin{equation}
    \begin{aligned}
        \sigma^2(\hat{x}_I)_J(t_J=0) &= \sigma^2(\hat{x}_J)_I(t_I=0) \\
        &\quad -\widetilde{\text{cov}}(\hat x_J,\hat v_{J\vert I}\hat t_J)_I(t_I=0) \\
        &\quad + \sigma^2(\hat{v}_{J\vert I}\hat{t}_J)_I(t_I=0).
        \label{eq:sigma(x_I)_J(0)}
    \end{aligned}
\end{equation}
Then, as illustrated in Fig. \ref{FIG:sigma_position}, the reciprocal variance between systems $I$ and $J$ is not the same. In particular, we note that the state of a clock in a given perspective may affect the variance of position and, hence, the spatial distribution of the latter in the perspective of the former. This aspect is one of our main results and will be further explored in Section~\ref{subsec:cssls}.

Before moving to the next section, we briefly discuss the conditions for the reduction of a spatiotemporal quantum reference frame(STQRF) into an SQRF. Intuitively, since the SQRF has a single ``classical'' notion of time, this reduction occurs when all clocks are synchronized and have zero relative uncertainty. With the tools introduced here, we can formalize this idea, as we do in Appendix~\ref{app:sqrf-limit}.

\section{Clock's relative uncertainty and the spatial distribution of systems} \label{sec:examples}

In this section, we illustrate two of the main results we have just derived. In particular, we show how the temporal parts of a system affect their spatial distribution upon a change of reference frame. This is the case even though, for simplicity, we assume no interactions with any of the temporal parts.

Consider a joint system of three subsystems ($A$, $B$, and $C$), each with spatial coordinates and individual clocks. Assume the initial state of the system in a given frame, say  $A$, is of the form $\Ket{\psi_{\bar A}(t_A=0)} = \ket{\xi_B} \otimes \ket{\lambda_B} \otimes \ket{\xi_C} \otimes \ket{\lambda_C}$, where $\ket{\xi_I} \coloneqq \int d\omega_I
\tilde{\xi}_I(\omega_I)\ket{\omega_I}$ and $\ket{\lambda_I} \coloneqq \int dp_I \tilde{\lambda}_I(p_I)\ket{p_I}$ for $I=B,C$. Then, using Eq.~\eqref{eq:psi-a-bar}, we identify
\begin{equation}
\begin{aligned}
    \psi_{\bar{A}}(t_A=0, \underline{\omega_{\bar{A}}}, \underline{p_{\bar{A}}}) &= 
    \Psi^{\text{kin}}(-H_{\bar{A}},\underline{\omega_{\bar{A}}}, -p_{\bar A},\underline{p_{\bar{A}}})\\
    &= \Pi_{I\neq A} \tilde{\xi}_I(\omega_I) \tilde{\lambda}_I(p_I).
    \label{eq:initial-state-ex}
\end{aligned}
\end{equation}
and
\begin{equation}
    \psi_{\bar{A}}(t_A, \underline{\omega_{\bar{A}}}, \underline{p_{\bar{A}}}) = e^{-i t_A H_{\bar{A}}} \Pi_{I\neq A} \tilde{\xi}_I(\omega_I) \tilde{\lambda}_I(p_I).
\end{equation}
Taking the Fourier transform of the last expression, we write
\begin{equation}
    \begin{aligned}
    \psi_{\bar A}(t_A, \underline{\omega_{\bar{A}}}, \underline{x_{\bar{A}}}) &=\frac{1}{2\pi} \int d\underline{p_{\bar{A}}} e^{i(x_{B}p_B+x_{C}p_C-t_AH_{\bar A})}\\ 
    &\hspace{15mm} \tilde{\xi}_B(\omega_B)\tilde{\xi}_C(\omega_C) \tilde{\lambda}_B(p_B)\tilde{\lambda}_C(p_C).
    \end{aligned}
    \label{eq:psi-a-bar_IC}
\end{equation}

Similarly, from Eqs.~\eqref{Psi_J}, \eqref{eq:psi-b-bar}, and~\eqref{eq:initial-state-ex}, we have
\begin{equation}
    \psi_{\bar B}(t_B, \underline{\omega_{\bar{B}}}, \underline{p_{\bar{B}}}) = e^{-i t_B H_{\bar{B}}} \tilde{\xi}_B(-H_{\bar{B}}) \tilde{\xi}_C(\omega_C) \tilde{\lambda}_B(-p_{\bar{B}}) \tilde{\lambda}_C(p_C),
\end{equation}
which has Fourier transform
\begin{equation}
    \begin{aligned}
        \psi_{\bar B}(t_B, \underline{\omega_{\bar{B}}}, \underline{x_{\bar{B}}}) &= \frac{1}{2\pi}\int d\underline{p_{\bar{B}}} e^{i(x_{A}p_A+x_{C}p_C-t_BH_{\bar B})}\\ 
        &\hspace{15mm} \tilde{\xi}_B(-H_{\bar B}) \tilde{\xi}_C(\om_C) \tilde{\lambda}_B(-p_{\bar B}) \tilde{\lambda}_C(p_C).
    \end{aligned}
    \label{eq:example-b-persp}
\end{equation}
This expression will be the basis for our analysis in this section.

\subsection{Frame dependency of entanglement}
\label{frame_dep_entngl}

The form of the initial conditions just discussed reveals that, while there was no entanglement between the systems from $A$'s perspective, particles $A$ and $B$ may be entangled from $C$'s perspective.

This can be seen from the factor $\tilde{\lambda}_B(-p_{\bar B})$ in Eq.~\eqref{eq:example-b-persp} since it suggests that the state of systems $A$ and $C$ will generally be mixed for an arbitrary $\lambda_B$. To give a concrete example, consider the special case in which $\xi_B$ and $\xi_C$ are constant functions, $\lambda_B(x_{B}) = 
e^{ik_B x_{B}}$, 
and $\lambda_C(x_{C}) = \delta(x_{C}-x_0)$, where $k_B$ and $x_0$ are real constants. Then, following the analysis presented at the beginning of this section, it can be checked that
\begin{equation}
    \psi_{\bar{B}}(t_B=0,\underline{\omega_{\bar B}}, \underline{x_{\bar B}})=
    e^{-ik_B{x_{A}}} \delta(x_{C} - x_{A} - x_0).
\end{equation}
Observe that this function can be equivalently written as 
$e^{-ik_B(x_{C}-x_0)}\delta(x_{C}-x_{A}-x_0)$. This means that $A$ and $C$ behave individually as plane waves. However, when one is measured, the other is localized a distance $x_0$ away from the location of the measured particle. 
In the above scenario, we see that particles that were not entangled in one frame become entangled in the other. This fits the spatial quantum reference frame treatment without using a dynamical time~\cite{angelo2011physics, Giacomini_2019}.

\subsection{Clock states and spatial localizability of systems}
\label{subsec:cssls}

We now turn our attention to the term $\xi_B(-H_{\bar{B}})$ in Eq.~\eqref{eq:example-b-persp}. The Hamiltonian $H_{\bar{B}}$ depends on spatial and temporal quantities of systems $A$ and $C$. As a result, depending on $\xi_B$ the various subsystems might be correlated in $B$'s perspective. Another way to put it is by saying that the state of a clock in a given perspective affects even the spatial distribution of particles in the perspective of the system to which that clock belongs.

To illustrate this, consider the case in which $\lambda_B(x_{B}) = (2/\pi\Delta^2)^{1/4} 
e^{-x_{B}^2/\Delta^2}
e^{iq_0x_{B}}$, $\lambda_C(x_{C}) = e^{ik_C {x_{C}}}$, and $\xi_I(\omega_I) \propto e^{-\alpha_I |\omega_I|}$, where $k_C,q_0 \in\mathbb{R}$, $\Delta>0$, $\alpha_I>0$, and $I=B,C$. According to Eq.~\eqref{eq:psi-a-bar_IC}, the wave function describing particles $B$ and $C$ from $A$'s perspective at an arbitrary instant of time $t_A$ is
\begin{widetext}
\begin{equation}
    \begin{aligned}
        \psi_{\bar A}(t_A,\underline{\omega_{\bar A}},\underline{x_{\bar A}}) \propto\; & e^{-\alpha_B|\omega_B|}e^{-\alpha_C|\omega_C|} \exp\left\{-it_A\left[\omega_B+\omega_C+\frac{(k_C+q_0)^2}{2m_A}+\frac{q_0^2}{2m_B} + \frac{k_C^2}{2m_C}\right]\right\} \\
        & {\sqrt{\frac{\pi}{\frac{\Delta^2}{4}+\frac{it_A}{2\mu_{BA}}}}} \exp\left\{-\frac{\left[x_{B}-t_A\left(\frac{k_C+q_0}{m_A}+\frac{q_0}{m_B}\right)\right]^2}{4\left(\frac{\Delta^2}{4}+\frac{it_A}{2\mu_{BA}}\right)}\right\} e^{ik_Cx_{C}+iq_0x_{B}},
    \end{aligned}
\end{equation}
where $\mu_{AB} = (1/m_A + 1/m_B)^{-1}$ is the reduced mass of particles $A$ and $B$. Using Eq.~\eqref{eq:example-b-persp}, the state of the system in $B$'s perspective can be obtained. We analyze here the resulting state for $\omega_A,\omega_C>0$. A similar analysis can be made for other relevant regions of the spectrum associated with $\omega_A$ and $\omega_C$. In the aforementioned domain, the wave function describing particles $A$ and $C$ from $B$'s perspective at an arbitrary instant of time $t_B$ is
\begin{equation}
    \begin{aligned}
        \psi_{\bar B}(t_B,\underline{\omega_{\bar B}},\underline{x_{\bar B}}) \propto\; & e^{-\alpha_C\om_C} \exp\left\{-\left(it_B+\alpha_B\right)\left(\om_A+\om_C+\frac{(k_C+q_0)^2}{2m_A}+\frac{q_0^2}{2m_B} + \frac{k_C^2}{2m_C}\right)\right\} \\
        & {\sqrt{\frac{\pi}{\frac{\De^2}{4}+\frac{\alpha_B}{2\mu_{AB}} + \frac{it_B}{2\mu_{AB}}}}} \exp\left\{-\frac{\left[x_{A}+(t_B-i\alpha_B) \left(\frac{k_C+q_0}{m_A}+\frac{q_0}{m_B}\right)\right]^2} {4\left(\frac{\De^2}{4}+\frac{\alpha_B}{2\mu_{AB}} + \frac{it_B}{2\mu_{AB}} \right)}\right\} e^{ik_Cx_{C}-i(q_0+k_C)x_{A}}.
    \end{aligned}
\end{equation}
\end{widetext}
Observe that even in the limit $\Delta \rightarrow 0$, when, in $A$'s perspective, particle $B$ is localized at $t_A=0$ [since $\lambda_B(x_{B}) \rightarrow \delta(x_B)$], the terms associated with the spatial distribution of particle $A$ from $B$'s perspective (i.e., the terms in the last line of the above equation) reveal that it has the form of a Gaussian for any $t_B$. This is due to the relative uncertainty of clock $B$ in $A$'s perspective. In fact, in the limit $\alpha_B\rightarrow0$, where the relative uncertainty between clocks $A$ and $B$ vanishes, it is possible to obtain a localized state for system $A$ at $t_B=0$ in $B$'s perspective. 
Similarly, we may choose a different normalized dependence of the clocks in $\omega_I$. Consider the case $\lambda_B(x_{B})=\delta(x_{B}-x_0)$, $\lambda_C(x_{C})=e^{ik_C x_{C}}$, and $\xi_I(\omega_I) = \text{rect}_{W_I}(\omega_I)$, where $x_0$, $k_C$, and $W_I$ are positive real constants, $I=B,C$, and
\begin{equation}
    \text{rect}_{W_I}\left(z\right)=
    \begin{cases}
        \frac{1}{W_I},\tab &-\frac{W_I}{2} < z < \frac{W_I}{2} \\
        0, \tab & \text{elsewhere}
    \end{cases}.
\end{equation}

Let $a = 1/2\mu_{AB}$, $b=k_C/m_B$, and $c = \omega_A + \omega_C + k_C^2/2\mu_{CB}$. If $-\frac{W_B}{2}<c-\frac{b^2}{4a}<\frac{W_B}{2}$, the initial state of particles $A$ and $C$ from $B$'s perspective is
\begin{equation}
    \begin{aligned}
        \psi_{\bar{B}}(t_B=0&,\underline{\omega_{\bar B}},\underline{x_{\bar B}}) = \\
        &\xi_C(\om_C) e^{ik_C(x_{C}+x_0)} 2e^{-i\frac{b}{2a}(x_{A} + x_0)}\\
        &\frac{\sin\left[p^{\text{eff}}(x_{A}+x_0)\right]}{p^{\text{eff}}(x_{A}+x_0)},
    \end{aligned}
\end{equation}
where $p^{\text{eff}}=\sqrt{\frac{b^2}{4a^2}-\frac{1}{a}\left(c-\frac{W_B}{2}\right)}$.

In the limit $W_B\ra\infty$, $p^{\text{eff}} \rightarrow \sqrt{W_B/2a}$, the result does not depend on any $\omega_I$. Precisely,
\begin{equation}
    \lim_{W_B\ra\infty}e^{-i\frac{b}{2a}(x_{A}+x_0)} \frac{\sin\left[p^{\text{eff}}(x_{A}+x_0)\right]}{p^{\text{eff}}(x_{A}+x_0)}\simeq \de(x_{A}+x_0).
\end{equation}
This is the case because, in this limit, the relative uncertainty between the clocks vanishes.

Observe that the results discussed here hold even though we do not apply any relativistic correction and, moreover, no interaction between the clocks and the spatial coordinates takes place. In fact, this can be seen as a consequence of the physical space not having the tensor product structure from the kinematical space, and the tensor product notation being used simply as a label for the systems~\cite{Trinity}. Moreover, because of the relative uncertainty between the clocks, the well-defined moment $t_A=0$ in $A$'s perspective becomes ``fuzzy'' in $B$'s perspective~\cite{Castro_Ruiz_2020}, hence the influence of the relative uncertainty between the clocks in the spatial distribution of systems.

\section{Discussion and outlook}
\label{sec:disc}

In this work, we have introduced a simple framework for 1+1 nonrelativistic spatiotemporal quantum reference frames, which combines position reference frames with the Page-Wootters approach. In our model, we have assumed each physical system contains, in addition to an external degree of freedom associated with its position, an internal degree of freedom that can be used as a clock. With this, each system can be used as a spatiotemporal reference frame and, moreover, the remaining systems satisfy the Schr\"odinger equation in the chosen perspective. 

In Section \ref{sec:ExpVal} we presented our main results, which consist of the derivation of formal expressions for perspective-dependent expectation values and variances associated with space, time, momentum, and the clock's energy degrees of freedom. We also presented relations between these quantities in different reference frames.

Recently, relativistic and nonrelativistic spatiotemporal reference frames have been considered in the literature~\cite{giacomini2021spacetime, Singh_2021, favalli2022model, giovannetti2023geometric}. While Ref.~\cite{favalli2022model}, like the present work, investigates a nonrelativistic framework, the authors consider a scenario with a single clock system to be used as a reference for time. One may think that this should indeed be enough for a nonrelativistic treatment, in particular in scenarios without interactions with the clock. However, as we have shown, the clock states in one perspective influence the spatial degrees of freedom of the systems when changing perspectives. For example, when only the spatial frame is considered, if a particle $B$ is localized in $A$'s perspective, then particle $A$ is also localized in $B$'s perspective \cite{2020_change_of_perspective}. However, in the spatiotemporal framework investigated here, this is not always the case. Depending on the state of clock $B$ in $A$'s perspective, particle $A$ might have a spatial spread in $B$'s reference frame even if $B$ is localized in $A$'s perspective, as shown in Section~\ref{sec:examples}. This happens despite the notion of time in every clock being the same in our study, as discussed in Section~\ref{sec:time}. In fact, it can be observed that the Heisenberg dynamics of the time operator $\hat{t}_J$ in system $I$'s perspective is
\begin{equation}
    \frac{d\hat{t}_J}{dt_I} = i [\hat{H}_{\bar{I}}, \hat{t}_J] = \mathds{1}_J
\end{equation}
for every $J\neq I$.

There are several directions that can be approached with the framework studied here. For instance, one could consider more general forms of the Hamiltonian $\hat{H}_T$ that include interactions between the different parts of the system, including the clocks. One could even consider a Hamiltonian whose interacting terms are such that the Hamiltonian and momentum constraints do not commute everywhere in the kinematical Hilbert space.

Furthermore, the results presented here can be extended to the 3+1 spatiotemporal scenario. In this case, the translational invariance constraint becomes effectively three individual constraints, one for each spatial direction. Additionally, one should also consider rotational invariance \cite{barbour2014identification, vanrietvelde2018switching}, which also introduces one constraint per rotational direction. Interestingly, let $\hat{x}_I^s$ be the spatial coordinate of system $I$ in the direction $s\in\mathfrak{R}\coloneqq\{x,y,z\}$ and $\hat{p}_I^s$ be its associated canonical momentum. Consequently, $\hat{P}_T^{r} = \sum_{K\in\mathfrak{I}} \hat{p}_K^{r}$ is the total momentum in the direction $r\in\mathfrak{R}$ and $\hat{L}_T^{u} = \sum_{J\in\mathfrak{I}} \hat{\ell}_J^{u}$ is the total angular momentum about the axis $u\in\mathfrak{R}$ with $\hat{\ell}_J^{u} = \sum_{v,s\in\mathfrak{R}} \epsilon_{uvs} \hat{x}_J^v \hat{p}_J^{s}$. Hence, it follows that
\begin{equation}
    \begin{aligned}
        [\hat{L}_T^{u}, \hat{P}_T^{r}] &= \sum_{J,K\in\mathfrak{I}} [\hat{\ell}_J^{u}, \hat{p}_K^{r}] \\
        &=\sum_{J,K\in\mathfrak{I}} 
        \sum_{v,s\in\mathfrak{R}} \epsilon_{uvs} [
        \hat{x}_J^v \hat{p}_J^{s}, \hat{p}_K^{r}] \\
        &= i \sum_{J\in\mathfrak{I}} \sum_{s\in\mathfrak{R}} \epsilon_{urs} \hat{p}_J^{s} 
        = i \sum_{s\in\mathfrak{R}} \epsilon_{urs} \hat{P}_T^{s},
    \end{aligned}
\end{equation}
where $\epsilon_{uvs}$ is the Levi-Civita symbol. Then, although these operators do not commute everywhere in the kinematical Hilbert space, they do commute everywhere in the hypersurface characterized by the translational invariance constraint. As a result, the rotational invariance constraint can be added, and the resulting physical Hilbert space coincides with the one obtained with the Hamiltonian and the total linear momentum constraints alone.

Finally, one could build the relativistic version of the framework we have considered here. In other words, by considering a system in which each subsystem contains external (i.e., spatial) degrees of freedom and an internal degree of freedom that serves as a clock, one should aim at modifying the constraints applied here in order to obtain the dynamics and frame transformations in the relativistic regime.

\acknowledgements{We thank Philipp H\"{o}hn for providing useful comments on an early version of this work. I.L.P. acknowledges support from the ERC Advanced Grant FLQuant. E.C. is supported by the Israeli Innovation Authority under Project 73795, by the Pazy Foundation, by the Israeli Ministry of Science and Technology, and by the Quantum Science and Technology Program of the Israeli Council of Higher Education.}

\bibliography{qrf_bib}

\onecolumngrid

\appendix

\section{Momentum and energy expectation values and covariances} \label{app:p-omega}

To derive the relation for the covariance between a pair of momenta in Eq.~\eqref{eq:sigma^2(p_I)_J}, we use Eqs.~\eqref{eq:<p_I>_any_short} and~\eqref{eq:<p_I_J=-sum} to write
\begin{equation}
    \braket{\hat{p}_I}_M\braket{\hat{p}_J}_M = -\braket{\hat{p}_I}_J\sum_{L\in\mathfrak{I}\setminus\{J\}} \braket{\hat{p}_L}_J = -\braket{\hat{p}_I}_J^2 -
    \sum_{L\in\mathfrak{I}\setminus\{I,J\}} \braket{\hat{p}_I}_J \braket{\hat{p}_L}_J.
\end{equation}
Moreover, we compute
\begin{equation}
    \begin{aligned}
        \braket{\hat{p}_I\hat{p}_J}_{M} &= \int d_{\bar{M}} \Psi_{\bar{M}}^* p_I p_J \Psi_{\bar{M}} = 
        \int d\underline{P}d\underline{\Omega} \delta(P_T)
        \delta(H_T)\Psi^* p_I p_J \Psi = -\int d_{\bar{J}}\Psi_{\bar{J}}^*p_I 
        \sum_{L\in\mathfrak{I}\setminus\{J\}} p_L \Psi_{\bar{J}} \\
        &= -\sum_{L\in\mathfrak{I}\setminus\{J\}}
        \braket{\hat{p}_I\hat{p}_L}_{J} = -\braket{\hat{p}_I^2}_{J} - 
        \sum_{L\in\mathfrak{I}\setminus\{I,J\}} \braket{\hat{p}_I\hat{p}_L}_J.
    \end{aligned}
    \label{eq_app:<p_I2>_bar_I}
\end{equation}
This allows us to write Eq.~\eqref{eq:sigma^2(p_I)_J}.

Now, to derive the covariance for a pair of clock energy operators in Eq.~\eqref{eq:sigma(omega_I)_J}, we first use the Wheeler-DeWitt constraint to rewrite the expectation value in Eq.~\eqref{eq:<omega_I>_J} as
\begin{equation}
    \braket{\hat{\omega}_I}_{J} = \int d_{\bar{J}} \Psi_{\bar{J}}^* \omega_I \Psi_{\bar{J}} = \int d_{\bar{I}} \Psi_{\bar{I}}^* \left(\omega_I\right)_{\bar{I}} \Psi_{\bar{I}} = -\int d_{\bar{I}} \Psi_{\bar{I}}^* \left(H_{\bar{I}}\right)_{\bar{I}} \Psi_{\bar{I}} = -\braket{\hat{H}_{\bar{I}}}_I = 
    -\sum_{L\in\mathfrak{I}\setminus\{ I\}} 
    \braket{\omega_L}_I - \braket{\mathcal{K}_{\bar{I}}}_I,
\end{equation}
With the above expression and Eq.~\eqref{eq:<omega_I>_J}, we write
\begin{equation}
    \braket{\hat{\omega}_I}_M\braket{\hat{\omega}_J}_M = -\braket{\hat{\omega}_I}_J \left(\sum_{L\in\mathfrak{I}\setminus\{ J\}} \braket{\hat{\omega}_L}_J + \braket{\hat{\mathcal{K}}_{\bar{J}}}_J\right) = -\braket{\hat{\omega}_I}_J^2 
    -\sum_{L\in\mathfrak{I}\setminus\{ I,J\}} \braket{\hat{\omega}_I}_J \braket{\hat{\omega}_L}_J -\braket{\hat{\omega}_I}_J \braket{\hat{\mathcal{K}}_{\bar{J}}}_J.
\end{equation}
Moreover, we compute
\begin{equation}
\begin{aligned}
\braket{\hat{\omega}_I\hat{\omega}_J}_M=
&\int d_{\bar M}\Psi_{\bar M}^* \omega_I \omega_J\Psi_{\bar M}=
\int d\underline{P}d\underline{\Omega}\delta(P_T)\delta(H_T)
\Psi^*\omega_I\omega_J\Psi=
\int d_{\bar J}\Psi_{\bar J}^* \omega_I 
\left(\sum_{L\in\mathfrak{I}\setminus\{ J\}}\omega_L+\mathcal{K}_{\bar J}\right)
\Psi_{\bar J}\\
=&-\sum_{L\in\mathfrak{I}\setminus\{ J\}} \braket{\hat{\omega}_I\hat{\omega}_L}_{J}-
\braket{\hat{\omega}_I\hat{\mathcal{K}}_{\bar J}}_J=
-\braket{\hat{\omega}_I^2}_{J}
-\sum_{L\in\mathfrak{I}\setminus\{ I,J\}} \braket{\hat{\omega}_I\hat{\omega}_L}_{J}-
\braket{\hat{\omega}_I\hat{\mathcal{K}}_{\bar J}}_J.
\end{aligned}
\end{equation}
With the last two expressions, we derive the relation in Eq.~\eqref{eq:sigma(omega_I)_J}.

\section{Time expectation values and variances} \label{Appendix:Time}

We start by deriving the variance of a time operator $\hat{t}_I$ in the reference frame of a system $J$ seen in Eq.~\eqref{eq:var-sigma-t}. For that, we compute
\begin{equation}
    \braket{\hat{t}_I^2}_{J}(t_J) = -\int d_{\bar{J}} \Psi_{\bar{J}}^* e^{i t_J H_{\bar{J}}} \frac{d^2}{d\omega_I^2}\left(e^{-i t_J H_{\bar J}} \Psi_{\bar{J}}\right) = \braket{\hat{t}_I^2}_J(t_J=0) + 2t_J \braket{\hat{t}_I}_J(t_J=0) + t_J^2.
\end{equation}
Together with Eq.~\eqref{<t_I>_J_short}, this leads to the expression for the variance in Eq.~\eqref{eq:var-sigma-t}.

Now, to derive the reciprocal relation for the expectation values in Eq.~\eqref{<t_I>_J(0)}, we observe that
\begin{equation}
    \begin{aligned}
        \frac{d}{d\omega_I} \Psi_{\bar{J}} &= \frac{d}{d\omega_I} \Psi^\text{kin}\left(\omega_J=-H_{\bar J},\underline{\omega_J},p_J=-p_{\bar J},\underline{p}_J\right) \\
        &= \left(\frac{\partial \Psi_{\bar J}}{\partial \omega_I} + \frac{\partial \Psi_{\bar J}}{\partial (-H_{\bar J})} \frac{\partial (-H_{\bar J})}{\partial \omega_I}\right) = \left[\left(\frac{d}{d\omega_I} - \frac{d}{d\omega_J}\right) \Psi\right]_{\bar J}.
    \end{aligned}
    \label{dd_omega_IPsi_J}
\end{equation}
In the last part of the above expression, the derivatives are taken in the kinematical Hilbert space (before the reduction to $J$'s perspective). This, together with Eq.~\eqref{eq:identity-any-frame}, allows us to write
\begin{equation}
    \left<f\left(\hat{t}_I\right)\right>_J(t_J=0) = \int d_{\bar I} \left[\Psi^* f\left(i\frac{d}{d\omega_I} - i\frac{d}{d\omega_J} \right) \Psi\right]_{\bar I} = \left<f\left(-\hat{t}_J\right)\right>_I(t_I=0).
    \label{eq:app:<f(t_I)>_J}
\end{equation}
for an arbitrary integrable real function $f$. In particular, if $f(x)=x$, we are lead to Eq.~\eqref{<t_I>_J(0)},
which we repeat here
\begin{equation}
    \braket{\hat{t}_I}_{J}(t_J=0) = -\braket{\hat{t}_J}_{I}(t_I=0).
    \label{app:<t_I>_J(0)}
\end{equation}
Moreover, if $f(x)=x^2$, we have
\begin{equation}
    \braket{t_I^2}_J(t_J=0)=\braket{t_J^2}_I(t_I=0),\label{app:<t_I^2>_J0=t_J^2>_I0}
\end{equation}
which, in turn, implies the reciprocal formula for the clocks' variance in Eq.~\eqref{eq:time-var-recip}.

\section{Position expectation values and variances}
\label{Appendix:Space}

To derive the expectation value of $\hat{x}_I$ in $J$'s reference frame in Eq.~\eqref{eq:x_I(t_J)}, we observe that
\begin{equation}
    \frac{d}{dp_I} H_{\bar{J}} = \frac{d}{dp_I}
    \left(\sum_{K\in\mathfrak{I}\setminus\{ J\}} \omega_K + \frac{p_{\bar{J}}^2}{2m_J} + 
    \sum_{K\in\mathfrak{I}\setminus\{ J\}} \frac{{p}_K^2}{2m_K}\right) = \frac{p_I}{m_I} + \frac{ p_{\bar{J}}}{m_J} = (v_{IJ})_{\bar{J}}
    \label{app:dH_J/dp_I}
\end{equation}
and
\begin{equation}
    \braket{\hat{x}_I}_J(t_J=0) = \int d_{\bar{J}} \Psi_{\bar{J}}^*i \frac{d}{dp_I}\Psi_{\bar{J}}.
    \label{<x_I>_J(0)}
\end{equation}
As a result,
\begin{equation}
    \begin{aligned}
        \braket{\hat x_I}_J(t_J) &= \int d_{\bar{J}} \Psi_{\bar{J}}^*e^{it_J H_{\bar{J}}} i\frac{d}{dp_I} e^{-it_J H_{\bar{J}}} \Psi_{\bar{J}} 
        = \int d_{\bar{J}} \Psi_{\bar{J}}^* \left[t_J \left(\frac{d}{dp_I}H_{\bar{J}}\right) + i\frac{d}{dp_I}\right] \Psi_{\bar{J}} \\
        &= \int d_{\bar{J}} \Psi_{\bar{J}}^* \left[\left(v_{IJ}\right)_{\bar{J}} t_J + i\frac{d}{dp_I}\right] \Psi_{\bar{J}} 
        = \braket{\hat{x}_I}_J(t_J=0) + t_J \braket{\hat{v}_{I\vert J}}_J,
        \label{app:<x_I>_J(t_J)}
    \end{aligned}
\end{equation}
which leads to Eq.~\eqref{eq:x_I(t_J)}.

Moreover, to derive the variance $\sigma^2(\hat{x}_I)_J$, we first compute
\begin{equation}
    \begin{aligned}
        \braket{\hat x_I^2}_J(t_J) &= \int d_{\bar{J}} \Psi_{\bar{J}}^* e^{-it_JH_{\bar{J}}} \left(i\frac{d}{dp_I}\right)^2 e^{it_JH_{\bar{J}}} \Psi_{\bar{J}} \\
        &= \int d_{\bar{J}} \Psi_{\bar{J}}^* \left[\left(i\frac{d}{dp_I}\right)^2 + t_J\left(i\frac{d}{dp_I} (v_{IJ})_{\bar{J}} + (v_{IJ})_{\bar{J}} i\frac{d}{dp_I}\right) + t_J^2(v_{IJ})_{\bar{J}}^2\right] \Psi_{\bar J} \\
        &= \braket{\hat{x}_I^2}_J(t_J=0) + \Big(\Braket{\hat{v}_{I\vert J} \hat{x}_I}_J(t_J=0) + t_J\braket{\hat{x}_I \hat{v}_{I\vert J}}_J(t_J=0)\Big) + t_J^2 \braket{\hat{v}_{I\vert J}^2}_J(t_J=0).
        \label{app:<x^2_I>J}
    \end{aligned}
\end{equation}
The last two expressions lead to Eq.~\eqref{eq:sigma(x_I)_J}.

Now, observe that
\begin{equation}
    \begin{aligned}
        \frac{d}{dp_I} \Psi_{\bar{J}} &= \frac{d}{dp_I} \Psi^\text{kin} \left(\omega_J=-H_{\bar{J}}, \underline{\omega_J}, p_J=-p_{\bar J}, \underline{p}_J\right) \\
        &= \frac{\partial \Psi_{\bar{J}}}{\partial p_I} + \frac{\partial \Psi_{\bar{J}}}{\partial (-p_{\bar{J}})} \frac{\partial (-p_{\bar{J}})}{\partial p_I} + \frac{\partial \Psi_{\bar{J}}}{\partial (-H_{\bar{J}})} \frac{\partial (-H_{\bar{J}})}{\partial p_I} \\
        &= \left[\left(\frac{d}{dp_I} - \frac{d}{dp_J} - v_{IJ}\frac{d}{d\omega_J}\right)\Psi\right]_{\bar{J}}.
    \end{aligned}
\end{equation}
Then, for an arbitrary real analytic function $f$, it holds that
\begin{equation}
    \begin{aligned}
        \left<f(\hat{x}_I)\right>_J(t_J=0) &= 
        \int d_{\bar{J}} \left[ \Psi^* f\left(i\frac{d}{dp_I} - i\frac{d}{dp_J} - v_{I\vert J} i\frac{d}{d\omega_J}\right)\Psi\right]_{\bar{J}} \\
        &= \int d_{\bar{I}} \left[ \Psi^* f\left(i\frac{d}{dp_I} - i\frac{d}{dp_J} +v_{J\vert I} i\frac{d}{d\omega_J}\right)\Psi\right]_{\bar{I}} \\
        &= \int d_{\bar{I}}
        \left[\Psi^* f\left(-\left(i\frac{d}{dp_J} - i\frac{d}{dp_I} - v_{J\vert I} i\frac{d}{d\omega_I}\right) + v_{J\vert I} i\left(\frac{d}{d\omega_J} - \frac{d}{d\omega_I}\right)\right)\Psi\right]_{\bar I} \\
        &= \braket{f(-\hat{x}_J + \hat{v}_{J\vert I} \hat{t}_J)}_I(t_I=0).
    \end{aligned}
    \label{eq:app:<f(x_I)>_J}
\end{equation}
With this, if $f(x)=x$, we obtain Eq.~\eqref{<x_I>J}. Moreover, letting $f(x)=x^2$, we have
\begin{equation}
    \braket{\hat{x}_I^2}_J(t_J=0) = \braket{\hat{x}_J^2}_I(t_I=0) - \braket{\left(\hat{x}_J\hat{v}_{J\vert I} + \hat{v}_{J\vert I} \hat{x}_J\right) \hat{t}_J}_I(t_I=0) + \braket{\hat{v}_{J\vert I}^2 \hat{t}_J^2}_I(t_I=0).
    \label{eq:app:<x_I^2>_J(0)}  
\end{equation}
Finally, from Eqs.~\eqref{<v_{I|J>_J}} and~\eqref{eq:app:<f(t_I)>_J}, we have
\begin{equation}
    \braket{\hat{v}_{I\vert J} \hat{t}_I}_J(t_J=0) =     \braket{\hat{v}_{J\vert I} \hat{t}_J}_I(t_I=0)
    \label{app:<vt>0}
\end{equation}
and
\begin{equation}
    \left<\hat{v}_{J\vert I}^2\hat{t}_J^2 \right>_I(t_I=0) = \left<\hat{v}_{I\vert J}^2\hat{t}_I^2 \right>_J(t_J=0).
\end{equation}
Using this and Eqs.~\eqref{app:<x_I>_J(t_J)} and~\eqref{eq:app:<x_I^2>_J(0)}, we are led to the relation between the reciprocal spatial variances at initial instances in Eq.~\eqref{eq:sigma(x_I)_J(0)}.

\section{The SQRF limit}
\label{app:sqrf-limit}

Repeating the steps in Section~\ref{sec:position} for SQRFs (described in Section \ref{subsec:SQRFs}), the expectation value of $\hat{x}_I$ in $J$'s reference frame, as function of $t$, is of the form
\begin{equation}
    \begin{aligned}
        \braket{\hat{x}_I}_J(t) &= \bra{\psi_{\bar{J}}(t)} \hat{x}_I \ket{\psi_{\bar{J}}(t)} \\
        &= \int d\underline{p_{\bar{J}}}
        \psi_{\bar{J}}^*(\underline{p_{\bar{J}}}) 
        e^{it H_{\bar{J}}} i\frac{d}{dp_I} e^{-itH_{\bar J}} \psi_{\bar J}(\underline{p_{\bar{J}}}) \\
        &=\int d\underline{p_{\bar{J}}} 
        \psi_{\bar{J}}^*(\underline{p_{\bar{J}}})
        \left(\left(v_{IJ}\right)_{\bar J}t + i\frac{d}{dp_I}\right)
        \psi_{\bar{J}}(\underline{p_{\bar{J}}}) \\
        &= \braket{\hat{x}_I}_J(t=0) + \braket{\hat{v}_{I \vert J}}_Jt,
    \end{aligned}
\end{equation}
where the expectation value at the initial instant of time is given by
\begin{equation}
    \braket{\hat{x}_I}_J(t=0) = \int d\underline{p_{\bar{J}}}
        \psi_{\bar{J}}^*(\underline{p_{\bar{J}}}) 
         i\frac{d}{dp_I} \psi_{\bar J}(\underline{p_{\bar{J}}}) =\int d\underline{p_{\bar{K}}} 
        \Big[\psi^*
        \left( i\partial_{p_I}-i\partial_{p_J}\right)
        \psi\Big]_{\bar{K}} = -\braket{\hat{x}_J}_I(t=0).
\end{equation}

Moreover, as expected, it can be seen that the reciprocal variances coincide, i.e., $\sigma^2(\hat{x}_I)_J(t)=\sigma^2(\hat{x}_J)_I(t)$, and the spatial description is mirror-opposite, i.e., $\braket{\hat{x}_I}_J(t)=-\braket{\hat{x}_J}_I(t)$. More generally, for an arbitrary analytic real function $f$ of a single variable, it holds that
\begin{equation}
    \braket{f(\hat{x}_I)}_J(t) = \braket{f(-\hat{x}_J)}_I(t).
    \label{<f(x_I)>_J_SQRF}
\end{equation}
The above expression contrasts with the analog expression in STQRFs, derived in Appendix~\ref{Appendix:Space}, which is
\begin{equation}
    \braket{f(\hat{x}_I)}_J(t_J=0) = \braket{f(-\hat{x}_J + \hat{v}_{J\vert I}\hat{t}_J)}_I(t_I=0).
    \label{eq:<f(x_I)>_J}
\end{equation}

From the last two expressions, we see that there are two ways for the STQRF to be reduced into a standard SQRF. The first, which is not physically interesting, requires all systems to have a fixed distance from one another (i.e., zero relative velocity). The other possibility coincides with the intuitive way to reduce to a SQRF: all clocks are synchronized and have zero relative uncertainty. Indeed, this follows from the condition that
\begin{equation}
    \hat{t}_J\ket{\psi_{\bar I}(0)} = 0
    \label{eq:t_Jpsi_I=0}
\end{equation}
for every $I\neq J$, which means that the states of all clocks in a given perspective are a Dirac delta distribution centered at zero. 

To conclude, we note that, although the state in Eq.~\eqref{eq:t_Jpsi_I=0} is non-normalizable, it is always possible to use a limit of normalizable wavepackets to have that behavior as an approximation.

\end{document}